%% file: main.tex
\documentclass[sigchi]{acmart}
\AtBeginDocument{%
  \providecommand\BibTeX{{%
    \normalfont B\kern-0.5em{\scshape i\kern-0.25em b}\kern-0.8em\TeX}}}

\newcommand{\quotedtext}[1]{%
   \noindent\textbf{\textit{{"#1"}}}%
}
\newcommand{\customheading}[1]{%
   \noindent\textbf{{#1}}%
}

\usepackage{enumitem}
\usepackage{ulem}
\usepackage{float}
\usepackage{soul}
\usepackage{xcolor}
\usepackage{titlesec}
\usepackage{subcaption}
\DeclareRobustCommand{\hlcyan}[1]{{\sethlcolor{white}\hl{#1}}}

\sethlcolor{white}
\setcopyright{acmcopyright}
\copyrightyear{2024}
\acmYear{2024}
\acmDOI{XXXXXXX.XXXXXXX}

\acmConference[CHI '24]{ACM (Association of Computing Machinery) CHI conference on Human Factors in Computing Systems}{May 11 - May 16, 2024}{Hawai'i, USA}
\acmPrice{15.00}
\acmISBN{978-1-4503-XXXX-X/18/06}

\begin{document}
%%
%% The "title" command has an optional parameter,
%% allowing the author to define a "short title" to be used in page headers.
\title[Evaluating ChatGPT's Usage and Impact in Indian Healthcare]{AI as a Medical Ally: Evaluating ChatGPT's Usage and Impact in Indian Healthcare}
% \titlenote{Jarvis : https://en.wikipedia.org/wiki/J.A.R.V.I.S.}

% \author{Ritvik Budhiraja$^{1}^*$, Ishika Joshi$^{2}^*$ \\ \normalsize $^{1}$IIITD Delhi \\ \normalsize e-mail: ritvik19322@iiitd.ac.in, ishika19310@iiitd.ac.in\\  }

% \thanks{*these authors contributed equally}

% \author{First Author\footnotemark[1]\and Second Author\footnotemark[1]\and Third Author}
% \renewcommand{\footnotetext}{\footnotetext{These authors contributed equally to this work.}}

\author{Aryaman Raina\textsuperscript{$ \dag  $}}
\email{aryaman20034@iiitd.ac.in}
\affiliation{%
  \institution{IIIT Delhi}
  % \streetaddress{Okhla Industrial Estate, Phase 3}
  \city{New Delhi}
  % \state{Delhi}
  \country{India}
  % \postcode{110020}
}

\author{Prateek Mishra\textsuperscript{$ \dag  $}}
\email{prateek20102@iiitd.ac.in}
\affiliation{%
  \institution{IIIT Delhi}
  % \streetaddress{Okhla Industrial Estate, Phase 3}
  \city{New Delhi}
  % \state{Delhi}
  \country{India}
  % \postcode{110020}
  }

\author{Harshit Goyal{$ \dag  $}}
\email{harshit20203@iiitd.ac.in}
\affiliation{%
  \institution{IIIT Delhi}
  % \streetaddress{Okhla Industrial Estate, Phase 3}
  \city{New Delhi}
  % \state{Delhi}
  \country{India}
  % \postcode{110020}
  }

\author{Dhruv Kumar}
\email{dhruv.kumar@iiitd.ac.in}
\affiliation{%
  \institution{IIIT Delhi}
  % \streetaddress{Okhla Industrial Estate, Phase 3}
  \city{New Delhi}
  % \state{Delhi}
  \country{India}
  % \postcode{110020}
}

% \thanks{*Equal Contribution} 
% \titlenote{\\
% \textsuperscript{$ \dag  $}Equal Contribution}

%%
%% By default, the full list of authors will be used in the page
%% headers. Often, this list is too long, and will overlap
%% other information printed in the page headers. This command allows
%% the author to define a more concise list
%% of authors' names for this purpose.
\renewcommand{\shortauthors}{Raina, Mishra, Goyal, et al.}

%%
%% The abstract is a short summary of the work to be presented in the
%% article.
\begin{abstract}

\input{files/00-abstract}
\end{abstract}

%%
%% The code below is generated by the tool at http://dl.acm.org/ccs.cfm.
%% Please copy and paste the code instead of the example below.
%%



%%
%% Keywords. The author(s) should pick words that accurately describe
%% the work being presented. Separate the keywords with commas.

%% A "teaser" image appears between the author and affiliation
%% information and the body of the document, and typically spans the
%% page.
% \begin{teaserfigure}
%   \includegraphics[width=\textwidth]{sampleteaser}
%   \caption{Seattle Mariners at Spring Training, 2010.}
%   \Description{Enjoying the baseball game from the third-base
%   seats. Ichiro Suzuki preparing to bat.}
%   \label{fig:teaser}
% \end{teaserfigure}

% \received{20 February 2007}
% \received[revised]{12 March 2009}
% \received[accepted]{5 June 2009}

%%
%% This command processes the author and affiliation and title
%% information and builds the first part of the formatted document.
\maketitle

\section{Introduction}
\input{files/01-introduction}

\section{Related Work}
\input{files/02-related_work}
\section{Methodology}
\input{files/03-methodology}

\section{Quantitative Evaluation}
\input{files/04x-Quantitative_Evaluation}
\section{Qualitative Evaluation}
\input{files/04-evaluation}

\vspace{-1em}
\section{Discussion}
\input{files/05-discussion}

% \vspace{-2em}
\section{Conclusion}

\input{files/06-conclusion}

% \section*{Acknowledgement}
% \begin{acks}
% \input{files/061-acknowledgements}
% \end{acks}
% \vspace{-1em}
% \section{Acknowledgments}
%%
%% The next two lines define the bibliography style to be used, and
%% the bibliography file.
\bibliographystyle{ACM-Reference-Format}
\bibliography{chatgpt-1}

%%
%% If your work has an appendix, this is the place to put it.
\appendix
\input{files/07-appendix}

\end{document}

%% file: files/00-abstract.tex
This study investigates the integration and impact of Large Language Models (LLMs), like ChatGPT, in India's healthcare sector. Our research employs a dual approach, engaging both general users and medical professionals through surveys and interviews respectively. Our findings reveal that healthcare professionals value ChatGPT in medical education and preliminary clinical settings, but exercise caution due to concerns about reliability, privacy, and the need for cross-verification with medical references. General users show a preference for AI interactions in healthcare, but concerns regarding accuracy and trust persist. The study underscores the need for these technologies to complement, not replace, human medical expertise, highlighting the importance of developing LLMs in collaboration with healthcare providers. This paper enhances the understanding of LLMs in healthcare, detailing current usage, user trust, and improvement areas. Our insights inform future research and development, underscoring the need for ethically compliant, user-focused LLM advancements that address healthcare-specific challenges.

%% file: files/01-introduction.tex
Large Language Models (LLMs), such as ChatGPT \cite{chatgpt} and Google Bard \cite{Google_Bard}, have recently emerged as transformative forces in the healthcare sector, redefining its future landscape. These advanced AI-driven systems, trained on vast datasets, have shown remarkable proficiency in various natural language processing tasks, including content creation, language translation, and code generation. Their integration into healthcare is not just a matter of technological advancement but a paradigm shift towards more efficient, patient-centric care systems.

% In the realm of healthcare, LLMs like ChatGPT offer promising opportunities for advancing healthcare delivery, particularly in improving and enhancing clinical decision support, analyzing unstructured and imaging data, creating documentation, summarizing research data, improving patient support and education, and many other business operations \cite{HIMSS2023}. Their “human-like” capabilities \cite{jamahealthforum} in generating comprehensive, intelligible text in response to complex inquiries can revolutionize patient-doctor communication, streamline medical documentation, and contribute to medical research through natural language interactions.

% These models have shown potential in areas such as drug discovery and cancer detection, adverse event detection, clinical documentation, and as virtual medical assistants. For instance, LLMs in the pharmaceutical landscape can detect adverse drug events and predict potential drug interactions \cite{aisera_llm_healthcare, medriva_llm_healthcare}. In oncology, they provide deeper insights into cancer progression by analyzing medical imaging datasets, leading to more targeted and effective treatment strategies \cite{medriva_llm_healthcare}. 

 In the realm of healthcare, LLMs like ChatGPT are transforming healthcare delivery by enhancing clinical decision support, analyzing diverse data types, and improving patient communication and education \cite{HIMSS2023}. These models are particularly effective in drug discovery, identifying adverse drug events, interpreting medical images for cancer detection, and serving as virtual medical assistants \cite{aisera_llm_healthcare}. Their ability to generate human-like text responses \cite{jamahealthforum} can revolutionize areas such as adverse event detection, clinical documentation, and medical research, offering significant advancements in fields like oncology \cite{medriva_llm_healthcare} and pharmaceuticals by facilitating more effective treatment strategies and predicting drug interactions. 

Our study seeks to explore the impact and usage of ChatGPT in healthcare comprehensively. We focus on how these models can be leveraged by medical professionals and general users, aiming to address the multifaceted challenges in the healthcare sector. Through comprehensive user studies, literature reviews, and interviews, our research aims to gather insights that will shape the development of virtual assistant solutions that meet the diverse needs of both expert and lay users. The objective is to harness the potential of LLMs, particularly ChatGPT, to revolutionize healthcare accessibility, medication management, and public health awareness. This research is pivotal in understanding how LLMs can be integrated into healthcare systems ethically and effectively, ensuring they complement rather than replace the human elements of patient care.

The paper employs a mixed-methods research approach \cite{mixed_methods}, drawing from both qualitative and quantitative data sources. This methodology includes the collection and analysis of \textbf{\textit{46 }}survey responses and the conduction of \textit{\textbf{6}} in-depth interviews with general users and medical professionals respectively, across India. The surveys and interviews are designed to address the following key research questions:
\begin{itemize}
\item \textbf{RQ1:} How are LLMs, particularly ChatGPT, being used by healthcare professionals and general users in India?
\item \textbf{RQ2:} What are the perceived benefits and challenges of using ChatGPT in healthcare settings?
\item \textbf{RQ3:} What insights and recommendations can be derived to enhance the integration of LLMs into healthcare systems?
\end{itemize}

The preliminary findings from our surveys and interviews reveal a diverse range of applications and perceptions. Among healthcare professionals, ChatGPT is predominantly utilized for tasks such as aiding in medical diagnosis, facilitating medical research, and enhancing patient communication. Meanwhile, general users primarily employ these models for gaining health information, understanding medical terminology, and seeking preliminary advice for health concerns. However, concerns regarding accuracy, ethical implications, and the need for human oversight are prevalent, emphasizing the necessity for a balanced integration of LLMs in healthcare.
%These findings suggest a growing acceptance of LLMs in healthcare, balanced by an awareness of their limitations and potential areas for further development. 

Moreover, our research extends to examining the role of ChatGPT in public health awareness and medication management. The insights gleaned from this study are not only significant for healthcare practitioners and patients but also carry implications for policymakers and technology developers. To the best of our knowledge, this study represents one of the first extensive explorations of the practical usage and impact of LLMs like ChatGPT in the healthcare sector in India. It is important to acknowledge that this is an ongoing investigation. To further solidify our understanding and ensure generalizability, we plan to conduct additional surveys and interviews with a broader range of participants. 
%The outcomes of this research are expected to contribute significantly to the evolving discourse on the integration of AI and LLMs in healthcare, providing valuable guidelines for future technological developments and implementation strategies

%% file: files/02-related_work.tex
The emergence of large language models (LLMs) like ChatGPT in healthcare has opened new frontiers in clinical practice. Parikh et al. \cite{parikh2023chatgpt} conducted an online survey revealing that healthcare professionals in India, though less likely to have used ChatGPT than their non-healthcare counterparts, generally hold a positive outlook towards its impact on their careers. Meanwhile, Mahajan et al. \cite{mahajan2019artificial} explored AI's transformative journey in healthcare, emphasizing its role in augmenting intelligence, especially in radiology, and addressing the challenges in developing nations like India. Notably, these studies primarily explore the anticipated effects of AI and ChatGPT on career trajectories in healthcare \cite{parikh2023chatgpt} and other sectors, rather than its actual utility and usage patterns in healthcare settings in India. In contrast, our research aims to explore the practical applications and utility of ChatGPT in the healthcare sector, examining how it influences and integrates into the everyday clinical and diagnostic process for general users and medical professionals.

Another study by Cascella et al. \cite{Cascella2023} highlighted ChatGPT's adeptness in structuring medical notes, showcasing its proficiency in accurately categorizing clinical parameters, even with minimal context. This ability to learn from errors and reassign misplaced parameters underpins its suitability in dynamic clinical environments where adaptability and precision are paramount. Additionally, in Saudi Arabian teleconsultations, healthcare professionals have leveraged ChatGPT for diagnostic assistance, notably in identifying symptoms and aiding in diagnosing novel infections \cite{Alanzi2023}. However, these studies also underscore ChatGPT's limitations, such as its lack of depth in medical expertise and challenges in understanding complex condition-treatment relationships, necessitating caution and human oversight in clinical applications \cite{Cascella2023, Alanzi2023}.

The adoption of ChatGPT in medical education has shown promising results. Hosseini et al. \cite{Hosseini2023} reported higher acceptability among medical trainees, particularly for administrative tasks, indicating a generational shift in the perception of AI tools. A notable application includes its use in United States Medical Licensing Examination (USMLE) \cite{USMLE2024} preparation, where ChatGPT demonstrated competence in posing relevant questions, indicating its potential as an educational aid \cite{Hosseini2023}. Other responses to ChatGPT’s role in education have been mixed, with some institutions encouraging its use and others imposing restrictions, reflecting the ongoing debate over AI's role in education \cite{Hosseini2023}.

In medical research settings, ChatGPT has shown promise in comprehending complex material. Cascella et al. \cite{Cascella2023} evaluated ChatGPT's capability in understanding and summarizing complex scientific materials from sources like the New England Journal of Medicine \cite{NEJM2024}. The model demonstrated accuracy in summarizing outcomes and identifying secondary findings but tended to exceed length constraints. Ethical concerns about the transparency of LLM training data and potential biases also emerged, highlighting the need for clarity in how these models are trained \cite{Hosseini2023}.

A comprehensive analysis by Zaman \cite{Zaman2023} provided a SWOT framework, detailing ChatGPT's strengths such as its vast knowledge repository and NLP capabilities, beneficial for administrative tasks and summarizing patient reports. Weaknesses included limitations in understanding nuances of patient language and lack of emotional intelligence. Opportunities were identified in patient engagement and diagnosis recommendations, while threats encompassed potential inaccuracies, biases, and ethical concerns around AI in clinical judgment \cite{Zaman2023}.

In triaging ophthalmic conditions, a study by Riley et al. \cite{Riley2023} compared ChatGPT 4's triage accuracy against human physicians. The model showed a high level of diagnostic accuracy, comparable to physicians, and outperformed other tools like Bing Chat \cite{bing} and the WebMD Symptom Checker \cite{WebMDSymptoms} in terms of accuracy and consistency.

Looking forward, the integration of LLMs in healthcare poses a spectrum of benefits and challenges as suggested by Gokul et al. \cite{Gokul2023}. While opportunities for enhanced medical learning, diagnostic accuracy, and global collaboration are evident, concerns regarding privacy, security, and ethical implications must be addressed to ensure responsible and secure usage of LLMs in healthcare applications.

%% file: files/03-methodology.tex
    \subsection{Research Design} 
    %% Yash Chillar
    We employed a mixed-methods \cite{mixed_methods} research design to examine the impact and usage of ChatGPT in the healthcare sector in India.  A mixed-methods design enabled the integration of qualitative and quantitative methods, offering a comprehensive understanding of the research matter. 
The study involved two primary research components: a survey targeting general users and in-depth interviews with healthcare professionals. These components were designed to capture diverse perspectives and provide insights into different aspects of ChatGPT usage in healthcare. The survey was conducted through Google Forms. This approach aimed to gauge the general user perspective on their usage of ChatGPT for healthcare purposes. Approximately 46 survey responses were collected, focusing on general attitudes, usage patterns, and perceived benefits and challenges of employing ChatGPT in everyday healthcare-related activities.

For the qualitative aspect of our study, we conducted in-depth interviews \cite{interaction_design_foundation_2022} with medical professionals, specifically medical students from various medical colleges across India. This selection was intentional to garner detailed insights from individuals who are both users and potential future implementers of LLMs in healthcare settings. A total of 6 interviews were conducted, with participants providing insights on their experiences, expectations, and professional viewpoints on the integration of LLMs like ChatGPT in their field. The interview participants were recruited through snowball sampling \cite{Snowball}. The interviews were conducted virtually via Google Meet \cite{GoogleMeet2024}, and the transcripts obtained from the audio recordings underwent additional analysis. Written and verbal consent were obtained from the interviewees before conducting the interviews and recording the sessions. 

The survey and interview questions underwent several rounds of revision and testing. An iterative design process \cite{IterativeDesign} was followed in refining the questions, enhancing the logical flow, and ensuring the elimination of ambiguities or biases. Pilot testing with a small group of users provided preliminary feedback, leading to further refinements.

    %% Sidhartha Garg
    \subsection{Data Collection and Analysis} 
    % A detailed data collection methodology was followed which utilised interviews and surveys to comprehensively examine the utilization, challenges, impact and perceptions of ChatGPT among undergraduate computer science students.

 \customheading{Survey:}
The survey, comprising four sections, began with an introduction highlighting its voluntary, confidential nature and research significance. It then collected demographic data such as age, gender, education, and experience with web technologies and LLMs to understand diverse user perspectives. The third section focused on the usage of web technologies in healthcare, examining patterns in seeking medical information, decision-making for doctor consultations, and information verification practices. The final section assessed the impact of AI tools like ChatGPT on healthcare decisions, exploring user motivations, satisfaction levels, challenges, and the influence on health choices, including preferences for text, voice, and visual AI interactions.
    
The survey utilized a mix of single-select, multi-select, and open-text response formats, allowing us to collect qualitative and quantitative data. Completion time was confined to an estimated 3-5 minutes to balance detailed feedback and respondent convenience. Survey dissemination was done through various channels, including private channels, university student email lists, and word-of-mouth. Responses were analyzed and utilized further to gain insights and in the framing of the interview questions for medical professionals. \hl{The complete list of questions asked in the survey is presented in Appendix} \ref{sec:survey_questionnaire}.
%     The survey comprised four main sections. The initial section of the survey served as an introduction, outlining the purpose and significance of the research while emphasizing the voluntary nature of participation and the confidentiality of responses.

% In the subsequent section, we gathered demographic information such as age, gender, education level, and the participants' experience with web technologies and LLMs. This demographic data was essential for understanding the diverse perspectives across different user groups. 

% The third section delved deeper into the specific usage of web technologies in healthcare. Questions in this section were designed to assess the frequency of medical information searches, types of web technologies utilized, online patterns for seeking medical information, decision-making processes regarding doctor consultations, and practices for verifying medical information. 

% In the fourth section, we specifically explored the impact of AI-powered tools like ChatGPT on healthcare decision-making. This section included questions about motivations for using AI tools, satisfaction with the information provided, challenges encountered, and the influence of these tools on health decisions. Options for text-based, voice-based, and visual interactions with AI tools were also examined to understand user preferences in interacting with AI technologies. 
    % The survey was circulated amongst university students using multiple channels including private channels, university student email lists, and word-of-mouth. 
    
    \customheading{Interviews:}
The interview questions for healthcare professionals centered on their interaction with technology, especially ChatGPT. Initial questions covered their roles and responsibilities in healthcare, followed by discussions on the integration of digital tools in the medical field. The focus then shifted to their information-seeking behaviors and decision-making in patient care, particularly the influence of external sources on patient treatment. Central to the interviews were detailed inquiries about their direct experiences with ChatGPT and similar AI tools, assessing its compatibility in healthcare settings, and any concerns about its integration into clinical practice. The final questions explored the necessary training for effective use of ChatGPT in healthcare, and solicited opinions on its limitations and potential ethical issues, aiming to understand the readiness and future implications of such technology in the healthcare sector. The interview format was semi-structured, allowing for flexibility and depth in responses while ensuring consistency across all interviews. The interviews were conducted virtually, ensuring convenience and accessibility for participants. The list of questions used is provided in Appendix \ref{sec:interview_questionnaire} of this paper.

    \customheading{Data Analysis:}  
    The audio recordings of the interviews were transcribed, followed by a Thematic Analysis (TA) \cite{braun_using_2006} on the overall collected data. Survey responses were systematically coded and organized into thematic categories. For the interview transcripts, a three-tier TA process was employed. This process began with semantic coding of transcripts \cite{braun_using_2006}, which were then combined into intermediate themes. Latent coding of these themes \cite{braun_using_2006} led to the final set of themes, forming the foundation of our study's findings and discussions.
% This process commenced with an initial semantic coding of the transcripts \cite{braun_using_2006}, followed by the amalgamation of these codes into intermediate themes. The final stage involved a latent coding of these intermediate themes \cite{braun_using_2006} to extract the definitive set of themes. These final themes from surveys and interviews formed the basis of our study's findings and discussions.  

   \subsection{Limitations}
% This study, while extensive in its scope and methodology, has limitations that warrant acknowledgment. The design of research tools was primarily exploratory, aimed at gaining initial insights into the use of ChatGPT in healthcare.   Due to the ongoing nature of the research, some aspects of the data collection and analysis are yet to be fully realized. This preliminary report represents an interim snapshot of a larger, evolving study.  
This study, though comprehensive, has limitations due to its exploratory design and ongoing nature. Initially focused on ChatGPT's use in healthcare, it presents preliminary findings from a larger, evolving research project. As part of our ongoing work, we plan to incorporate interviews with general users to get their deeper insights.
% The study predominantly relied on qualitative methods (in-depth interviews) with medical professionals and quantitative methods (surveys) with general users. The use of surveys for general users was a deliberate choice aimed at leveraging their accessibility to capture a broader range of perspectives. In contrast, the choice to conduct in-depth interviews with medical professionals, primarily medical students, was informed by our access to a specific, though limited, set of professionals, which may have led to unintentional bias. We recognize this as a limitation and are actively working to expand our network to include a larger and more varied group of medical professionals. Additionally, a more symmetrical application of qualitative and quantitative methods across all participant categories and age groups could potentially yield a more nuanced understanding of the diverse perspectives and experiences. 
Our study primarily used qualitative interviews with medical professionals, mainly medical students, and quantitative surveys with general users. While surveys captured a broad range of user perspectives, interviews were limited to a specific group of professionals, possibly introducing bias. We acknowledge this limitation and are working to involve a more diverse group of medical professionals in future research. A balanced use of qualitative and quantitative methods across all participant categories could provide a deeper understanding of varying perspectives. Moreover, this study was conducted in India's cultural and geographical context. Therefore, the findings might not be directly applicable to other regions or healthcare systems with different cultural, economic, and technological environments.

%% file: files/04x-Quantitative_Evaluation.tex
% \subsection{Demographics and Evaluation of Survey-responses}

Using the survey formulated for the study, we were able to collect a total of 46 responses from general users. Visualizations of the key findings can be found in Tables~\ref{table:web_technologies} to~\ref{table:interaction_preferences} (for multi-select questions) and Figures~\ref{fig:age_group} to~\ref{fig:verify_ai_info} (for single-select questions) in the Appendix. This section will provide a summary of the results, focusing on the most significant findings due to space constraints. 

The survey's demographic section revealed that a significant majority of the respondents, 80.4\%, fell within the 18-24 age group, followed by 13\% in the 40-64 age range, 6.5\% in the 25-39 category, and none above 65 years. Gender distribution was predominantly male, with 73.9\%, while females represented 23.9\% of the respondents. 2.2\% of respondents identified as non-binary. Education levels varied among participants: 43.5\% were high school graduates or equivalent, 34.8\% held a bachelor’s degree, 17.4\% had a master’s degree or higher, and 4.3\% possessed an associate or vocational degree. Regarding experience with web technologies and LLMs like ChatGPT, 50\% reported moderate experience, 17.4\% had extensive experience, 26.1\% had limited experience, and 6.5\% had no experience at all.

When asked about the usage of web technologies for seeking information related to disease diagnosis or medical treatments, 28.3\% of participants indicated a frequency rating of 4 out of 5, with 30.4\% scoring 3, 19.6\% scoring 2, 15.2\% scoring 5, and 6.5\% never using these technologies for this purpose. The most commonly used web technologies for gathering medical information were search engines like Google and Bing (91.3\%), followed by medical websites such as WebMD and Mayo Clinic (47.8\%), ChatGPT and similar AI tools (37\%), online medical forums and communities (21.7\%), and health-related apps (10.9\%).

In terms of online information-seeking behavior, researching specific symptoms was most common (89.1\%), understanding treatment options came next (69.6\%), reading about personal experiences (43.5\%), seeking second opinions (26.1\%), comparing different sources (32.6\%), and other unspecified methods (2.2\%). Regarding the timing of consulting a doctor, 50\% of respondents preferred to do so after self-research and initial home remedies, followed by 21.7\% only when the condition seems serious, 15.2\% when symptoms worsen, 10.9\% immediately after noticing symptoms, and 2.2\% under other unspecified circumstances. 43.5\% of participants occasionally verified a doctor's diagnosis or treatment recommendation by searching for information online, 23.9\% did so often, and 32.6\% never did so.

Exploring the use of AI tools for healthcare information, the main motivations for using these tools included quick answers to health-related questions (63\%), convenience and accessibility (47.8\%), seeking information outside of regular office hours (23.9\%), privacy concerns (23.9\%), second opinions (23.9\%), and other reasons (4.4\%). The respondents rated the accuracy and relevance of AI tools’ information as follows: 37\% rated it 3 out of 5, 28.3\% rated it 4, 21.7\% rated it 2, 8.7\% rated it 5, and 4.3\% rated it 1.

% Challenges encountered in using AI tools for healthcare information included uncertainty about the source of information (58.7\%), concerns about information accuracy (52.2\%), lack of personalized advice (34.8\%), ambiguity in responses (21.7\%), difficulty understanding medical jargon (23.9\%), and other reasons (6.6\%). Regarding verification of information provided by AI tools, 52.2\% did so occasionally, 30.4\% often, and 17.4\% never. In terms of the influence of AI tools on health decisions, 54.3\% of respondents reported no influence, 26.1\% were unsure, and 19.6\% acknowledged some influence. Preferred methods of interacting with AI tools for healthcare purposes included text-based chat (82.6\%), visual representations like graphs or diagrams (41.3\%), and voice-based interaction (28.3\%).

In using AI tools for healthcare, challenges included uncertainty about information sources (58.7\%), accuracy concerns (52.2\%), lack of personalized advice (34.8\%), response ambiguity (21.7\%), and difficulty with medical jargon (23.9\%), with other issues at 6.6\%. For verifying AI-provided information, 52.2\% did so occasionally, 30.4\% often, and 17.4\% never. AI's influence on health decisions was reported as none by 54.3\%, uncertain by 26.1\%, and some influence by 19.6\%. Preferred interaction methods with AI were text-based chat (82.6\%), visual aids like graphs (41.3\%), and voice interaction (28.3\%). 

%% file: files/04-evaluation.tex
Utilizing a three-layer thematic analysis \cite{braun_using_2006}, we analysed the qualitative data from interviews and surveys. This data was compiled and sorted into three main themes: \textit{Usage Patterns and Benefits, Challenges, and Perceptions and Recommendations}.
% \newcommand{\quotedtext}[1]{%
%    \noindent\textbf{\textit{{"#1"}}}%
% }

% Table \ref{tab:interview_demographics} depicts the demographics of the participants of the interviews. The participants were taken from 8 different universities. Out of 17 participants, 8 were male and 9 were female. 6 participants were freshmen, 5 participants were sophomores, 2 participants were juniors and 3 were seniors. This representation is influenced by the varying availability of students across different academic years owing to commitments such as internship drive preparations and placement preparations.
% \subsection{Emerging Usage Patterns and Benefits of using ChatGPT}
\subsection{Usage Patterns and Benefits of ChatGPT}
\customheading{Usage Patterns adopted by Medical Students and Practitioners.} The thematic analysis of the interview transcripts reveals several patterns in the usage of ChatGPT in the healthcare sector, highlighting its benefits and applications. Medical students and practitioners have explored ChatGPT for various purposes, ranging from medicinal academic assistance to preliminary diagnosis aid. 

Medical students frequently utilize ChatGPT as an educational tool. It assists them in understanding complex medical cases, summarizing information, and aiding in new research. Students mentioned using ChatGPT to "summarize" academic material and to get a "rough vague idea" of certain medical concepts. For instance, a student described using ChatGPT for “solving questions like if we have been given a medical case”.

\quotedtext{\hlcyan{ChatGPT has been very useful for my research. [...] For instance, I will type out what are the new areas of research [...] give me an idea for a new research paper. [..] I found that pretty cool.}}

Some participants explored the use of ChatGPT for preliminary diagnostics, although with caution. They perceived it as a tool that could potentially assist in diagnosing simple medical cases or providing a starting point for further investigation. However, they emphasized the need for professional medical confirmation, suggesting ChatGPT as a supplementary tool rather than a standalone diagnostic solution.

\quotedtext{\hlcyan{But I must say that, you know, like, it is okay to use it in the majority of the cases, but you know, the risk factor is higher or there is a severity in the case [...] like in emergency situations.}}

A recurring benefit highlighted was the efficiency and time-saving aspect of using ChatGPT. Participants found it particularly useful for quickly accessing medical information, which is crucial in the fast-paced environment of medical studies and practice. They underscore the valuable role of ChatGPT in managing the extensive information requirements in medical education and practice, providing quick access to necessary data and helping manage time more effectively.

\quotedtext{\hlcyan{And also, like I got the specific answer I wanted. Like, that's the difference I guess is from simple Google search [...]. But when I ask the same question from ChatGPT, then the specific answer to what I needed was provided.}}

% \quotedtext{\hlcyan{We just open the app and type the query and we get our answer. So, I like this case.}
% }

\subsection{Challenges with the use of ChatGPT}
A significant theme emerging from the interviews is the concern regarding the limitations and reliability of ChatGPT in making accurate diagnoses. Participants expressed skepticism about relying solely on ChatGPT for diagnostic purposes, especially in critical cases. A key challenge identified is the model's inability to provide specific, detailed answers required for precise medical diagnosis. This sentiment was echoed in another interview, where the participant highlighted the need for further verification of ChatGPT's outputs.

\quotedtext{I think the ChatGPT answers are pretty vague. [...] if I ask it a certain question like what is the cutoff value for anaemia or things like that, the answers given by ChatGPT are generally a bit vague. They're not so specific.}

Another major issue raised in the interviews concerns the ethical implications and privacy concerns associated with using ChatGPT in healthcare. Participants were wary about the handling of sensitive personal health information, emphasizing the need for robust data protection measures.  One interviewee pointed out the potential risk of personal health information being accessed by unauthorized parties:

\quotedtext{When we're typing such personal information [...] that could potentially be leaked somewhere and it could cause a huge ethical concern.}
In addition to privacy issues, there is also a concern about the ethical use of AI in healthcare, particularly in situations where the AI's recommendations could have significant health implications.

The interviews also highlighted challenges in terms of usability and accessibility of ChatGPT for diverse users, especially those from non-technical or rural backgrounds. Participants pointed out that the current mode of interaction with ChatGPT, primarily through typing, might not be feasible for everyone, particularly in rural areas where digital literacy is lower. One participant suggested improvements to make ChatGPT more accessible. The recommendation was to develop features that allow for more natural interaction, such as voice commands and possibly even interpreting expressions or non-verbal cues: 

\quotedtext{Communicating with AI should be made more natural like using people's voice, expressions [...] for people living in rural areas [...] able to communicate through speech and expressions would be more suitable.}
\subsection{Perceptions and Recommendations}

Participants expressed the need for ChatGPT to integrate standard medical textbooks and reference materials to enhance its reliability and usefulness. A medical student emphasized the importance of sourcing information from recognized medical texts, suggesting that direct references to these sources in ChatGPT's responses would be beneficial.

\quotedtext{\hlcyan{So to improve it, I feel we need to integrate more of these standard textbooks [...] whenever it gives me an answer, it should specifically point out where it took the answer from.}} 

There was a strong sentiment among interviewees about the necessity of establishing legal frameworks and standards for the use of AI tools like ChatGPT in healthcare. These regulations should focus on ensuring patient privacy, accuracy of medical information, and preventing the misuse of AI for critical medical decision-making. The interviewees suggested that such regulations could be crafted and enforced by health authorities or government bodies.

\quotedtext{\hlcyan{The legal regulation should [...] make sure that the companies don't leak any data, they don't sell any data regarding the healthcare information of the people.}} 

Participants also suggested the customization of ChatGPT to cater to specific medical contexts and needs. This could include variations of the tool tailored for different specialities within medicine, such as paediatrics, oncology, or general practice. They suggested that such specialization could enhance the accuracy and relevance of ChatGPT's responses, making it a more useful tool for practitioners in those fields.

\quotedtext{\hlcyan{It would be nice if you were able to just copy-paste x-rays or scans [...] for instance, a patient has an MRI taken.}}

%% file: files/05-discussion.tex
Our research uncovers a nuanced understanding of the integration and impact of ChatGPT in the healthcare sector, particularly in India. The study reveals diverse usage patterns, acknowledges the challenges, and suggests insightful recommendations for the future integration of LLMs like ChatGPT in healthcare. Despite the emergence of Large Language Models fine-tuned for medical queries, such as Med-PaLM \cite{Singhal2023}, these specialized models are not as ubiquitous in day-to-day user interactions. Therefore, our study concentrates on the impact and usage of more readily available and adopted LLMs like ChatGPT. Our rationale lies in their broad availability and ease of use for the general public, making ChatGPT more representative of the current general user interaction with AI in healthcare contexts.   
% These findings offer a solid foundation for health technology developers, medical practitioners, and policymakers to harness the benefits of LLMs while addressing the existing limitations.

% Our survey highlights that users primarily seek AI tools due to their quick health-related answers, convenience and accessibility. However, this reliance is tempered by concerns about the accuracy and source of information, with many users uncertain about the credibility of the advice received. This dichotomy suggests a trust in AI for initial health inquiries, yet a hesitation to fully rely on these tools for serious health decisions, as indicated by a majority reporting little to no influence of AI tools for final health decisions. 

% The prevalent use of ChatGPT by medical students and professionals for educational purposes and preliminary diagnostics is notable. This trend aligns with global shifts towards AI-assisted learning and decision-making in healthcare. However, the reliance on ChatGPT for initial diagnostics, albeit cautiously, underlines the need for rigorous validation of such tools in clinical settings. Similar to existing diagnostic aids, ChatGPT could be fine-tuned and tested for specific medical scenarios, enhancing its reliability and utility in clinical practice. This approach demands a collaborative effort involving technologists, clinicians, and regulatory bodies to ensure the safe and effective deployment of ChatGPT in healthcare settings.

Our survey results indicate that while users appreciate the speed, convenience, and accessibility of AI tools like ChatGPT for health-related inquiries, there's a marked hesitation to depend on these for serious health decisions due to concerns about their accuracy and source credibility, as indicated by a majority reporting little to no influence of AI tools for final health decisions. In the medical field, both students and professionals are increasingly turning to ChatGPT for education and preliminary diagnostics, reflecting a global trend towards AI integration in healthcare. This cautious reliance on ChatGPT for initial diagnostics underscores the necessity for its rigorous validation in clinical environments. The potential for ChatGPT as a reliable clinical tool demands collaborative efforts from technologists, clinicians, and regulators, ensuring its safe and effective application in healthcare.

% Our study also highlights significant apprehensions regarding the reliability and ethical implications of using ChatGPT in healthcare. These concerns mirror the broader discourse on AI ethics and accountability in sensitive sectors like healthcare. Addressing these issues necessitates the development of transparent AI models that are explainable and auditable, ensuring that healthcare professionals can trust and understand the logic behind ChatGPT's responses. Additionally, stringent data protection and privacy measures must be established and adhered to, considering the sensitive nature of healthcare data.

Our study also highlights significant apprehensions regarding the reliability and ethical implications of using ChatGPT in healthcare, reflecting the wider issues in AI ethics and accountability.  Addressing these issues necessitates the development of transparent and explainable AI models that healthcare professionals can trust and understand. Moreover, rigorous data protection and privacy protocols are crucial due to the sensitive nature of healthcare data. 

ChatGPT holds great potential in enhancing patient-doctor communication and patient education, with future versions possibly offering more empathetic and accurate responses, along with personalized health advice. This could improve patient experiences and reduce healthcare professionals' workloads. Adapting ChatGPT for specific medical contexts, including voice and non-verbal interactions, is especially beneficial in diverse, multilingual environments like India, where digital literacy levels vary widely. Developing versions tailored to various medical fields and patient groups could make ChatGPT in healthcare more inclusive and user-friendly. 

% ChatGPT's potential in transforming patient-doctor communication and patient education is immense. Future versions could be developed to understand and respond to patient queries more empathetically and accurately, offering personalized health advice and support. Such advancements would improve patient experience and alleviate the workload on healthcare professionals. The suggestion to tailor ChatGPT for specific medical contexts and needs, including provisions for voice-based and non-verbal interactions can significantly enhance accessibility, especially in diverse and multilingual settings like India, where digital literacy levels vary widely. Developers should consider building versions of ChatGPT that are specialized for different medical disciplines and patient demographics, thereby making AI in healthcare more inclusive and user-friendly.

%% file: files/06-conclusion.tex
The study explores the integration of ChatGPT into India's healthcare sector, revealing its growing use among medical professionals for education, preliminary diagnostics, and research. While its efficiency in providing medical information is recognized, concerns about accuracy, ethical implications, and the need for professional verification are noted. General users show a preference for AI in healthcare, yet remain cautious about information trustworthiness and source credibility.

Key findings emphasize the importance of balancing ChatGPT's ethical integration in healthcare, highlighting the need for transparent AI models and strong data protection, especially due to the sensitive nature of healthcare data. The study suggests ChatGPT should augment, not replace, human medical expertise, and points to its potential in improving patient-doctor communication and education. Future development should focus on empathetic, accurate responses and tailoring for specific medical contexts, including voice-based and non-verbal interactions, to enhance accessibility in diverse settings like India. Our future research will expand on these findings, employing a balanced approach of qualitative and quantitative methods at a larger scale to explore ChatGPT's diverse healthcare applications.

%% file: files/07-appendix.tex
\onecolumn
\vspace{-1em}
\clearpage
\section{Survey Questionnaire}\label{sec:survey_questionnaire}
Questions with standard bullets ($\bullet$) are single-select while questions with square ($\tiny\square$) are multi-select.

\textbf{Section 1: Introduction} \
(Description of the survey's purpose and consent agreement)

\textbf{Section 2: Background Information}

\begin{enumerate}
\item Age Group (Please select one)

% %\begin{itemize}
$\bullet$ \ \textnormal{18-24} \ \ \
$\bullet$ \ \textnormal{25-39} \ \ \ 
$\bullet$ \ \textnormal{40-64} \ \ \ 
$\bullet$ \ \textnormal{65 or above}
% %%\end{itemize}
\item Gender (Please select one)

% %\begin{itemize}
$\bullet$ \ \textnormal{Male} \ \ \ 
$\bullet$ \ \textnormal{Female}\ \ \ 
$\bullet$ \ \textnormal{Non-binary}
% %\end{itemize}
\item Education Level (Please select the highest level completed)

% %\begin{itemize}
$\bullet$ \ \textnormal{High school graduate or equivalent} \ \ \ \ \
$\bullet$ \ \textnormal{Associate/Vocational degree} \ \ \ \ \ 
$\bullet$ \ \textnormal{Bachelor’s degree} \ \ \ \ \
$\bullet$ \ \textnormal{Master’s degree or Higher}
% %\end{itemize}
\item Experience with Web Technologies and LLMs like ChatGPT (Please select one)

% %\begin{itemize}
$\bullet$ \ \textnormal{No experience} \ \ \ \ \
$\bullet$ \ \textnormal{Limited experience (Use sometimes, basic understanding)} \ \ \ \ \
\\
$\bullet$ \ \textnormal{Moderate experience (Use regularly, good understanding)} \ \ \ \ \
$\bullet$ \ \textnormal{Extensive experience (Highly experienced and knowledgeable)}
% %\end{itemize}
\end{enumerate}

\textbf{Section 3: Web Technologies and Healthcare}

\begin{enumerate}
\setcounter{enumi}{4}
\item How often do you use web technologies (websites, apps, etc.) to search for information related to disease diagnosis or medical treatments?

%%\begin{itemize}
$\bullet$ \ \textnormal{1 (Very Rarely)} \ \ \
$\bullet$ \ \textnormal{2} \ \ \
$\bullet$ \ \textnormal{3} \ \ \
$\bullet$ \ \textnormal{4} \ \ \
$\bullet$ \ \textnormal{5 (Very Often)}
%\end{itemize}
\renewcommand{\labelitemi}{\tiny$\square$}
\item Which web technologies do you commonly use for gathering medical information? (Select all that apply)
\begin{itemize}
\item Search Engines (Google, Bing, etc.)
\item Medical websites (WebMD, Mayo Clinic, etc.)
\item Health-related apps
\item Online medical forums and communities
\item ChatGPT, Bard, Bing AI, or such tools
\item Other [Text field]
\end{itemize}
\item What patterns do you follow when seeking medical information online? (Select all that apply)
\begin{itemize}
\item Researching specific symptoms
\item Understanding treatment options
\item Reading about personal experiences
\item Comparing different sources
\item Seeking second opinions
\item Other [Text field]
\end{itemize}
\renewcommand{\labelitemi}{$\bullet$}
\item When do you generally decide to consult a doctor for a health concern?
\begin{itemize}
\item Immediately after noticing symptoms
\item After self-research and initial home remedies
\item When symptoms worsen
\item Only when the condition seems serious
\item Other [Text field]
\end{itemize}
\item Do you ever verify a doctor's diagnosis or treatment recommendation by searching for information online?

%\begin{itemize}
$\bullet$ \ \textnormal{Yes, often} \ \ \
$\bullet$ \ \textnormal{Yes, occasionally} \ \ \
$\bullet$ \ \textnormal{No, never}
%%\end{itemize}
\end{enumerate}

\textbf{Section 4: Exploring the Impact of AI-Powered Tools on Healthcare Decision-Making}

\begin{enumerate}
\setcounter{enumi}{9}
\renewcommand{\labelitemi}{\tiny$\square$}
\item If you've used AI Tools (like ChatGPT) for healthcare purposes, what motivated you to do so? (Select all that apply)
\begin{itemize}
\item Convenience and accessibility
\item Quick answers to health-related questions
\item Seeking information outside of regular office hours
\item Privacy (no need to share personal details with real people)
\item Second opinions
\item Other [Text field]
\end{itemize}
\renewcommand{\labelitemi}{$\bullet$}
\item How would you rate the accuracy and relevance of the information provided by AI Tools (like ChatGPT) for your healthcare queries?

%\begin{itemize}
$\bullet$ \ \textnormal{1 (Not accurate and relevant at all)} \ \
$\bullet$ \ \textnormal{2} \ \
$\bullet$ \ \textnormal{3} \ \
$\bullet$ \ \textnormal{4} \ \
$\bullet$ \ \textnormal{5 (Extremely accurate and relevant)}
%%\end{itemize}
\item Please provide examples of the types of healthcare questions you have used AI Tools (like ChatGPT) to answer. Were you satisfied with the responses you received? [Optional open-ended question. Participants were allowed to write without any word limit.]
\renewcommand{\labelitemi}{\tiny$\square$}
\item What challenges have you encountered while using AI Tools (like ChatGPT) for healthcare information? (Select all that apply)
\begin{itemize}
\item Difficulty in understanding medical jargon
\item Ambiguity in responses
\item Lack of personalized advice
\item Concerns about information accuracy
\item Uncertainty about the source of information
\item Other [Text field]
\end{itemize}
\renewcommand{\labelitemi}{$\bullet$}
\item Do you take any steps to verify the information provided by AI Tools (like ChatGPT) using other sources, such as consulting a medical professional or cross-referencing with reputable websites?

%\begin{itemize}
$\bullet$ \ \textnormal{Yes, often} \ \ \
$\bullet$ \ \textnormal{Yes, occasionally} \ \ \
$\bullet$ \ \textnormal{No, never}
%%\end{itemize}
\item Has using AI Tools (like ChatGPT) influenced any decisions you've made about your health? (e.g., seeking medical help, changing lifestyle habits, etc.)

%\begin{itemize}
$\bullet$ \ \textnormal{Yes} \ \ \
$\bullet$ \ \textnormal{No} \ \ \
$\bullet$ \ \textnormal{Not sure}
%%\end{itemize}
\renewcommand{\labelitemi}{\tiny$\square$}
\item How do you prefer to interact with AI Tools (like ChatGPT) for healthcare purposes? (Select all that apply)
\begin{itemize}
\item Text-based chat
\item Voice-based interaction
\item Visual representations (graphs, diagrams, etc.)
\item Other [Text field]
\end{itemize}
\item Is there anything else you would like to share about your experiences with ChatGPT in healthcare? [Optional open-ended question. Participants were allowed to write without any word limit.]
\end{enumerate}
\vspace{-1em}
% \section{Interview Questionnaire}\label{sec:interview_questionnaire}
% Below, we present the questionnaire template utilized by our research team during student interviews. It should be noted that this list does not cover all the questions, as the research team frequently posed additional inquiries in response to the participants' answers.
% \begin{enumerate}
%     \item How long have you been using ChatGPT for? How did you come across ChatGPT? How often do you use ChatGPT?
%     \item What have you been using ChatGPT for? Do you have any specific use cases? Have you used ChatGPT before to learn new concepts?
%     \item How do you think ChatGPT has supported your workflow, more specifically in your coursework?
%     \item Have you run into problems with ChatGPT? Do you come across scenarios that might be undesirable or obstructing? How do you tackle such issues? How do you deal with cases that are not answered correctly by ChatGPT?
%     \item What is your overall experience been like with ChatGPT? What is your current opinion of ChatGPT and the role it could play in changing the form of education? Threat or assistance?
%     \item Are there any suggestions that you would like to incorporate in ChatGPT in order to support your computer science related workflow better?
% \end{enumerate}
\section{Interview Questionnaire}\label{sec:interview_questionnaire}
Below, we present the questionnaire template utilized during interviews with medical professionals. It should be noted that this list does not cover all the questions, as the research team frequently posed additional inquiries in response to the participants' answers.

\textbf{Background and Practice:}
\begin{enumerate}
\item Can you briefly describe your current role and responsibilities within the healthcare domain?
\item How has the landscape of healthcare changed during your tenure, especially in terms of technology integration?
\end{enumerate}

\textbf{Information Seeking and Clinical Decision Making:}
\begin{enumerate}
\setcounter{enumi}{2}
\item How often do you feel the need to consult external sources for patient care, and what are these sources typically?
\item Can you recall a recent instance where an external source greatly aided in patient care or clinical decision-making?
\end{enumerate}

\textbf{Experience and Perspective on ChatGPT:}
\begin{enumerate}
\setcounter{enumi}{4}
\item Are you currently using, or have you ever used, Large Language Models or similar AI tools in your practice? Can you describe that experience?
\item How do you think ChatGPT might fit into the current healthcare ecosystem, especially in aiding professionals like yourself?
\item What are your primary concerns or reservations about incorporating ChatGPT into regular clinical practice?
\end{enumerate}

\textbf{Training and Implementation:}
\begin{enumerate}
\setcounter{enumi}{7}
\item How do you envision the ideal training or orientation for healthcare professionals before integrating ChatGPT into their practice?
\item In your opinion, are there specific areas within healthcare where ChatGPT should not be applied? Why?
\end{enumerate}

\clearpage

\section{Graphs and Tables for Quantitative Results}
\begin{figure*}
    \centering
    % First Row
    \begin{subfigure}{0.45\textwidth}
        \includegraphics[width=0.95\textwidth]{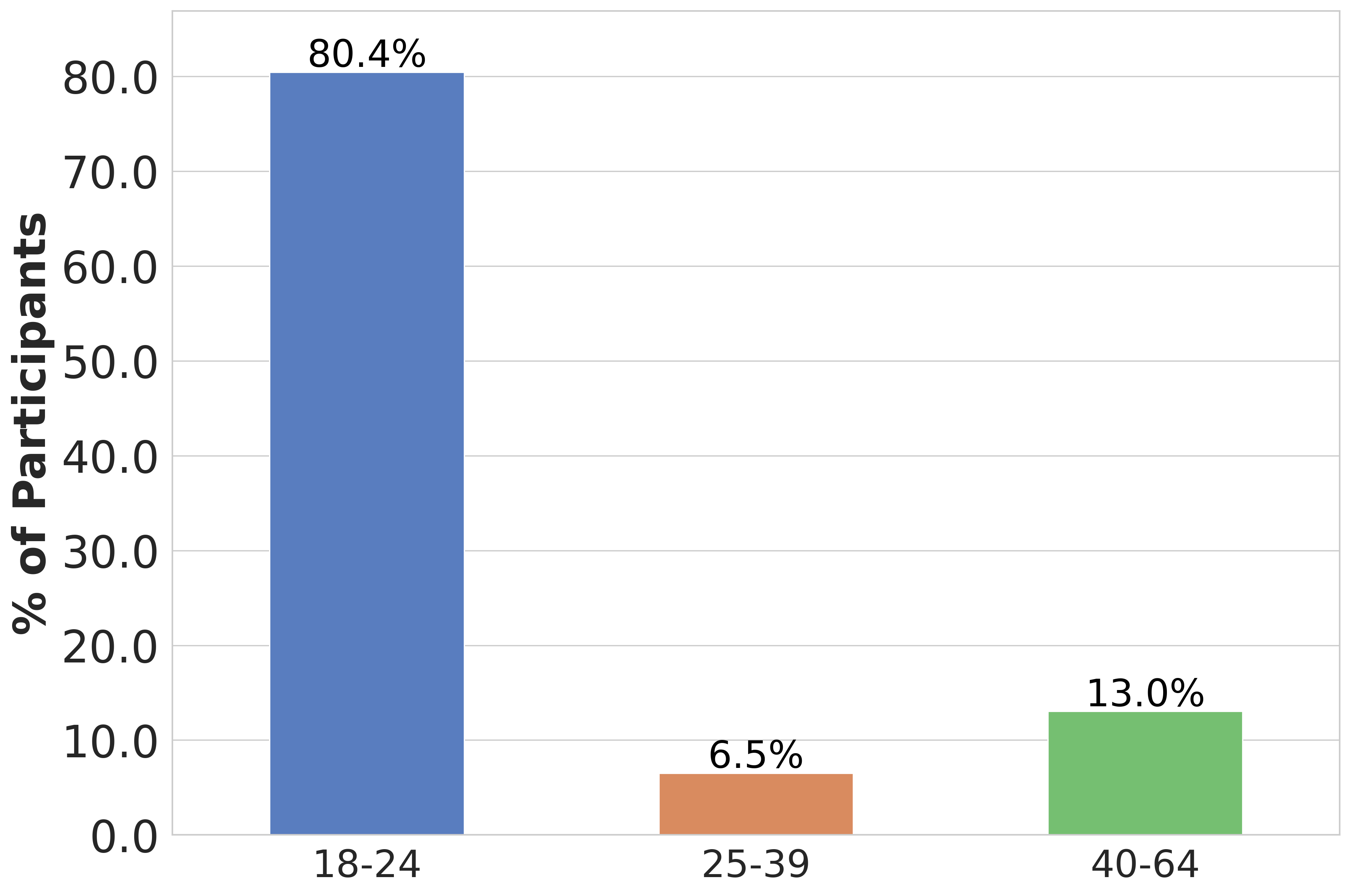}
        \caption{Age Group distribution of survey participants}
        \label{fig:age_group}
    \end{subfigure}
    \hfill
    \begin{subfigure}{0.45\textwidth}
        \includegraphics[width=0.95\textwidth]{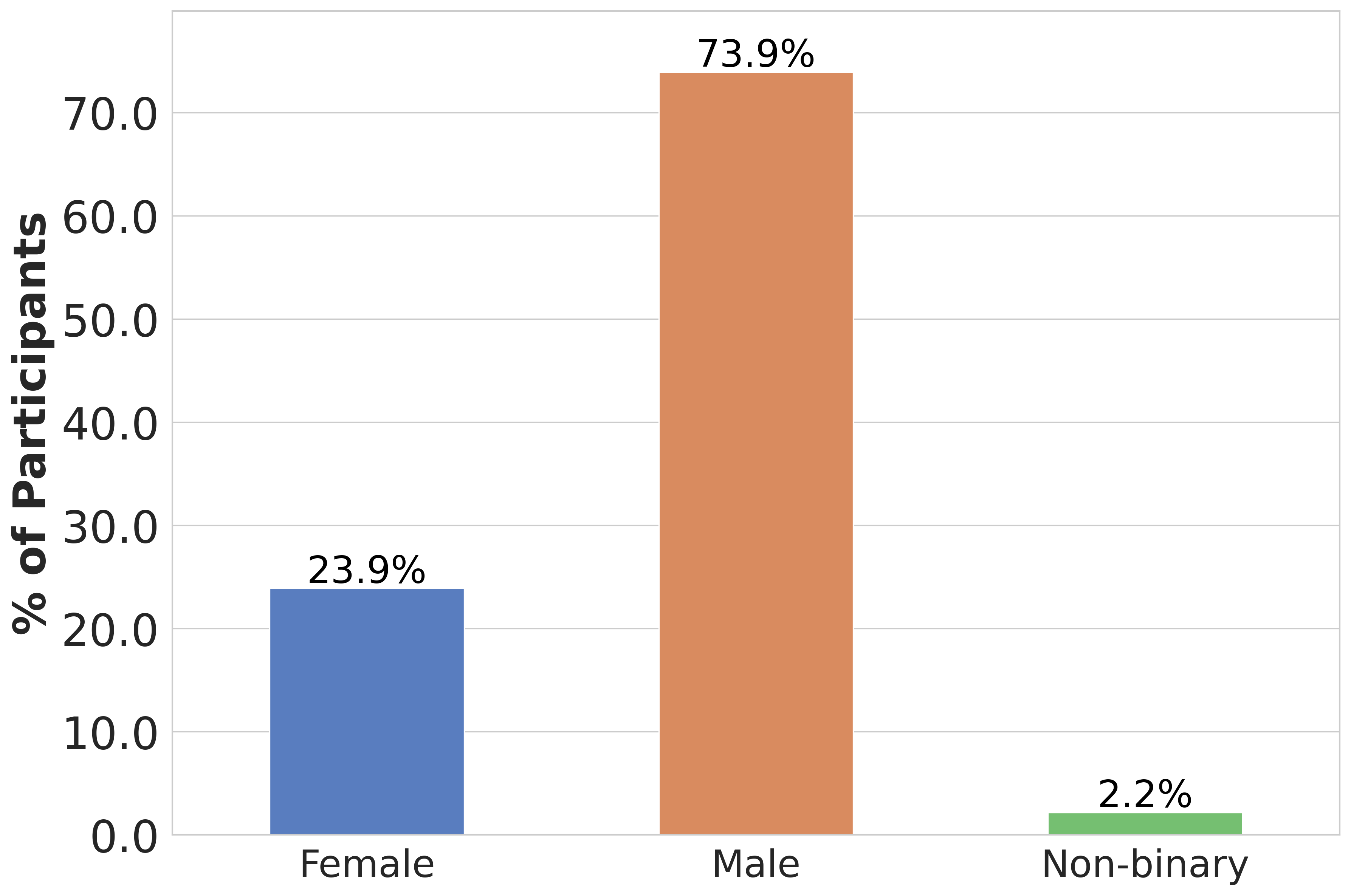}
        \caption{Gender distribution of survey participants}
        \label{fig:gender}
    \end{subfigure}

    % Second Row
    \begin{subfigure}{0.45\textwidth}
        \includegraphics[width=0.95\textwidth]{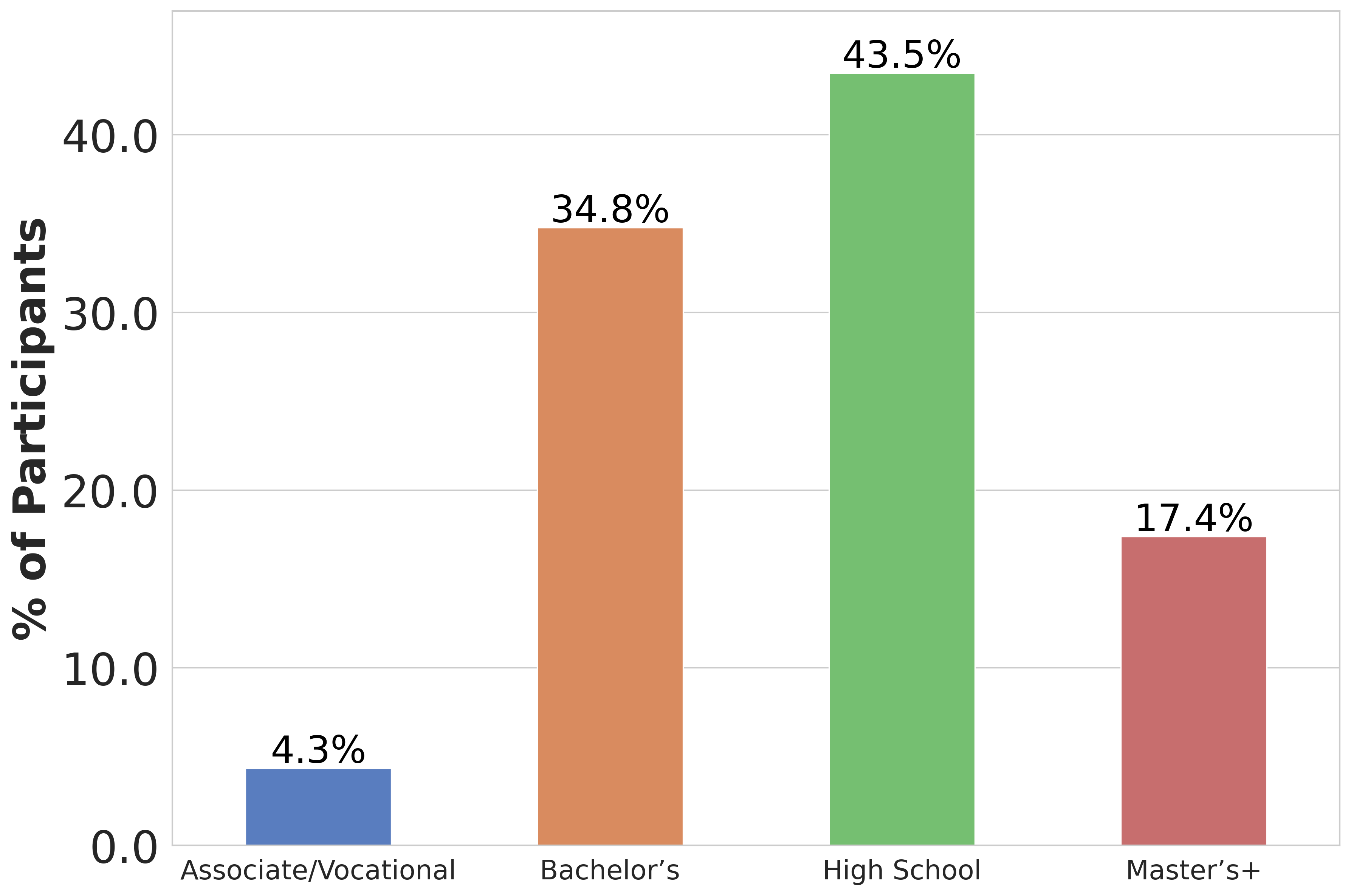}
        \caption{Education Level distribution of survey participants}
        \label{fig:education_level}
    \end{subfigure}
    \hfill
    \begin{subfigure}{0.45\textwidth}
        \includegraphics[width=0.95\textwidth]{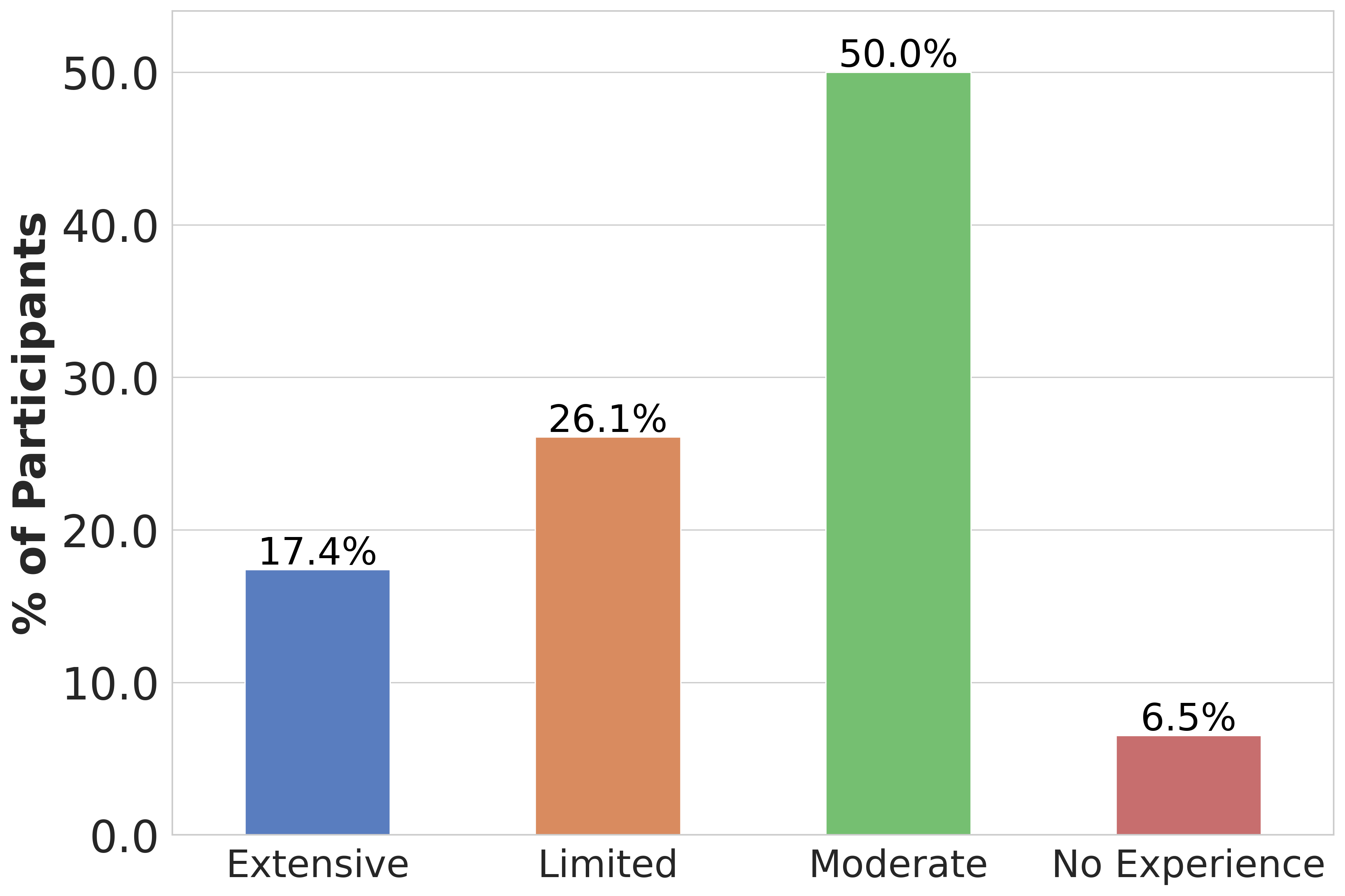}
        \caption{Experience with Web Technologies and LLMs like ChatGPT}
        \label{fig:exp_web_tech}
    \end{subfigure}

    % Third Row
    \begin{subfigure}{0.45\textwidth}
        \includegraphics[width=0.95\textwidth]{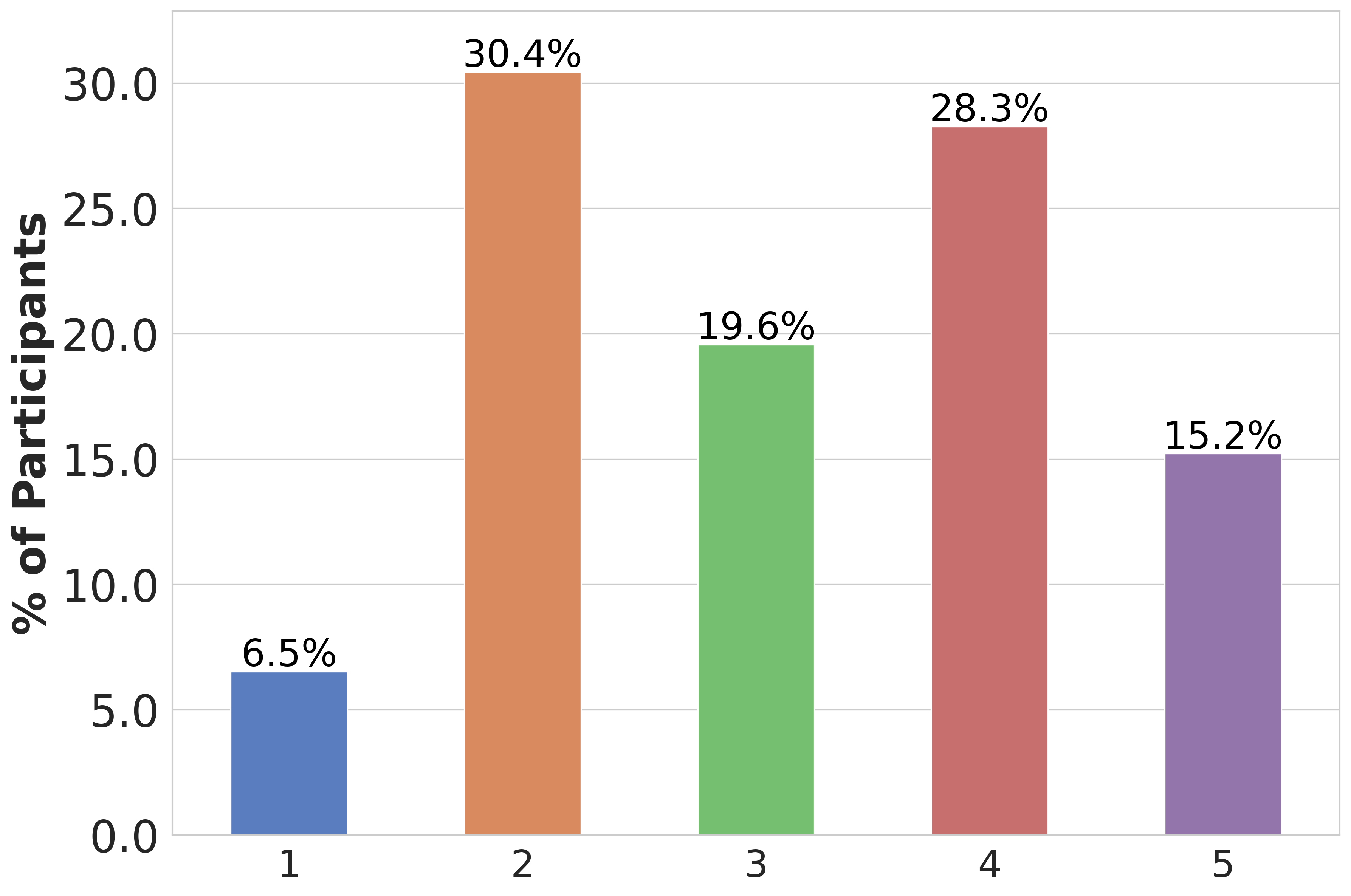}
        \caption{Frequency of Using Web Technologies for Medical Information}
        \label{fig:freq_web_tech_med_info}
    \end{subfigure}
    \hfill
    \begin{subfigure}{0.45\textwidth}
        \includegraphics[width=0.95\textwidth]{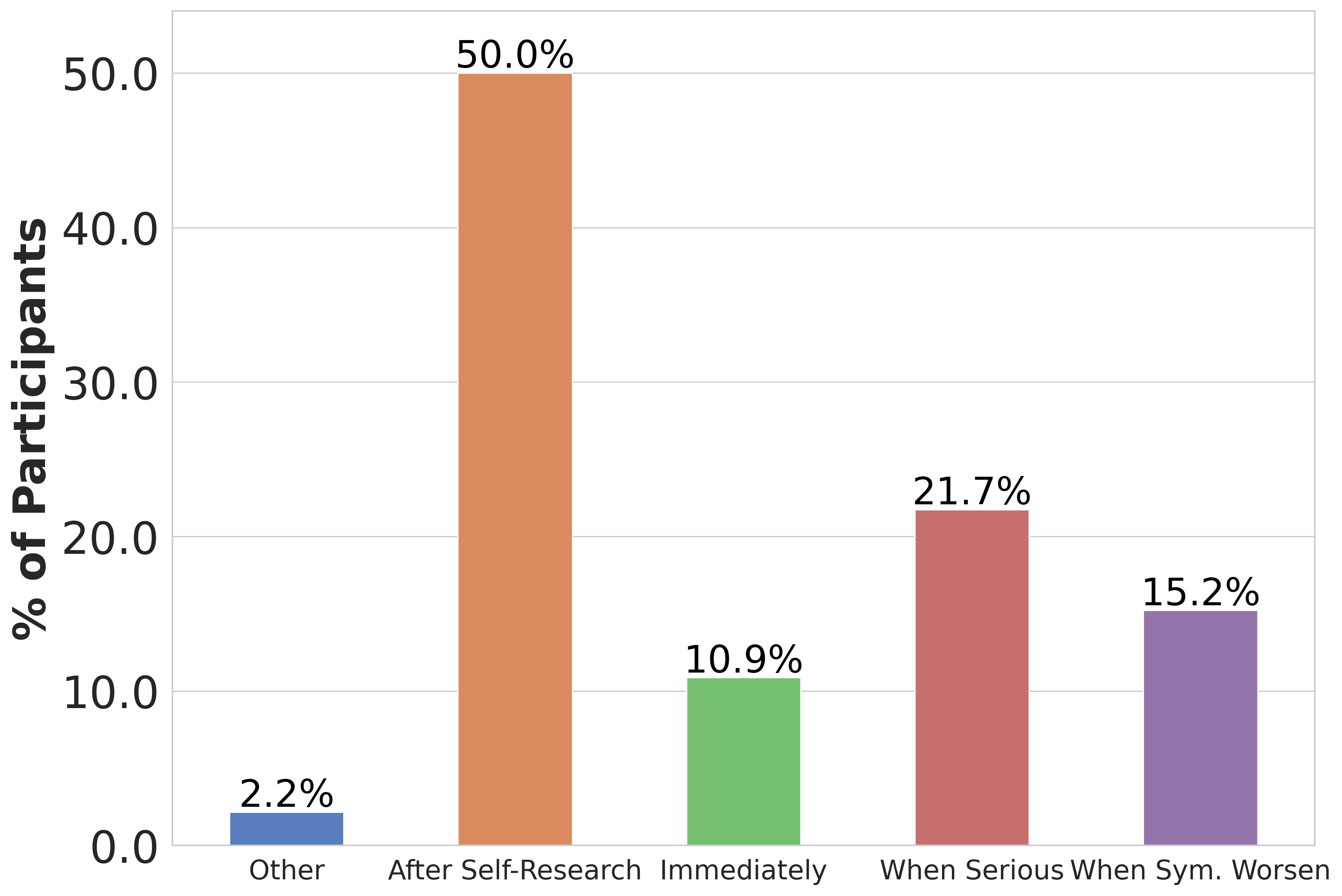}
        \caption{Decision to Consult a Doctor}
        \label{fig:consult_doctor}
    \end{subfigure}

    % Fourth Row
    \begin{subfigure}{0.45\textwidth}
        \includegraphics[width=0.95\textwidth]{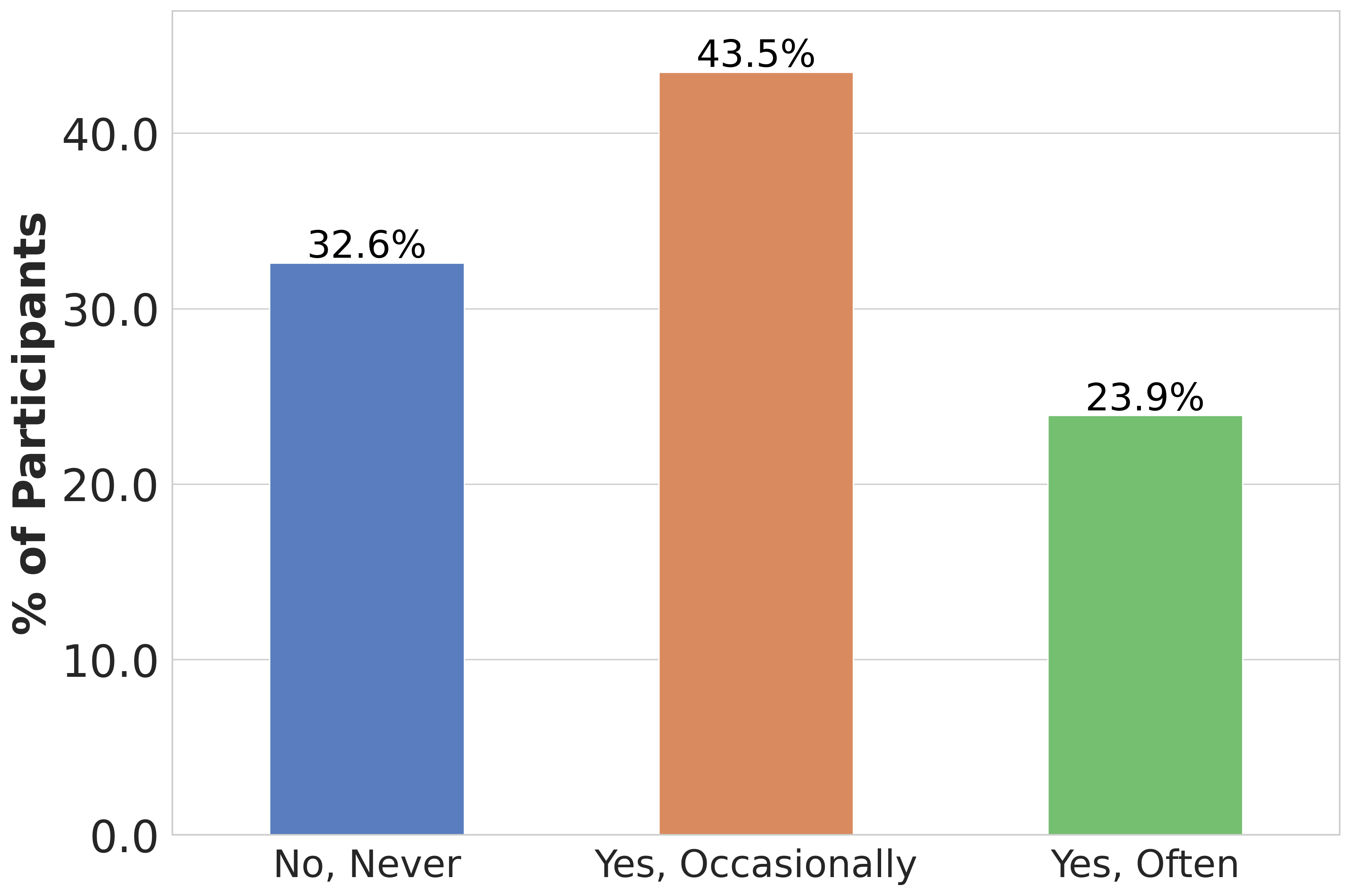}
        \caption{Verifying Doctor’s Diagnosis Online}
        \label{fig:verify_doctor_online}
    \end{subfigure}
    \hfill
    \begin{subfigure}{0.45\textwidth}
        \includegraphics[width=0.95\textwidth]{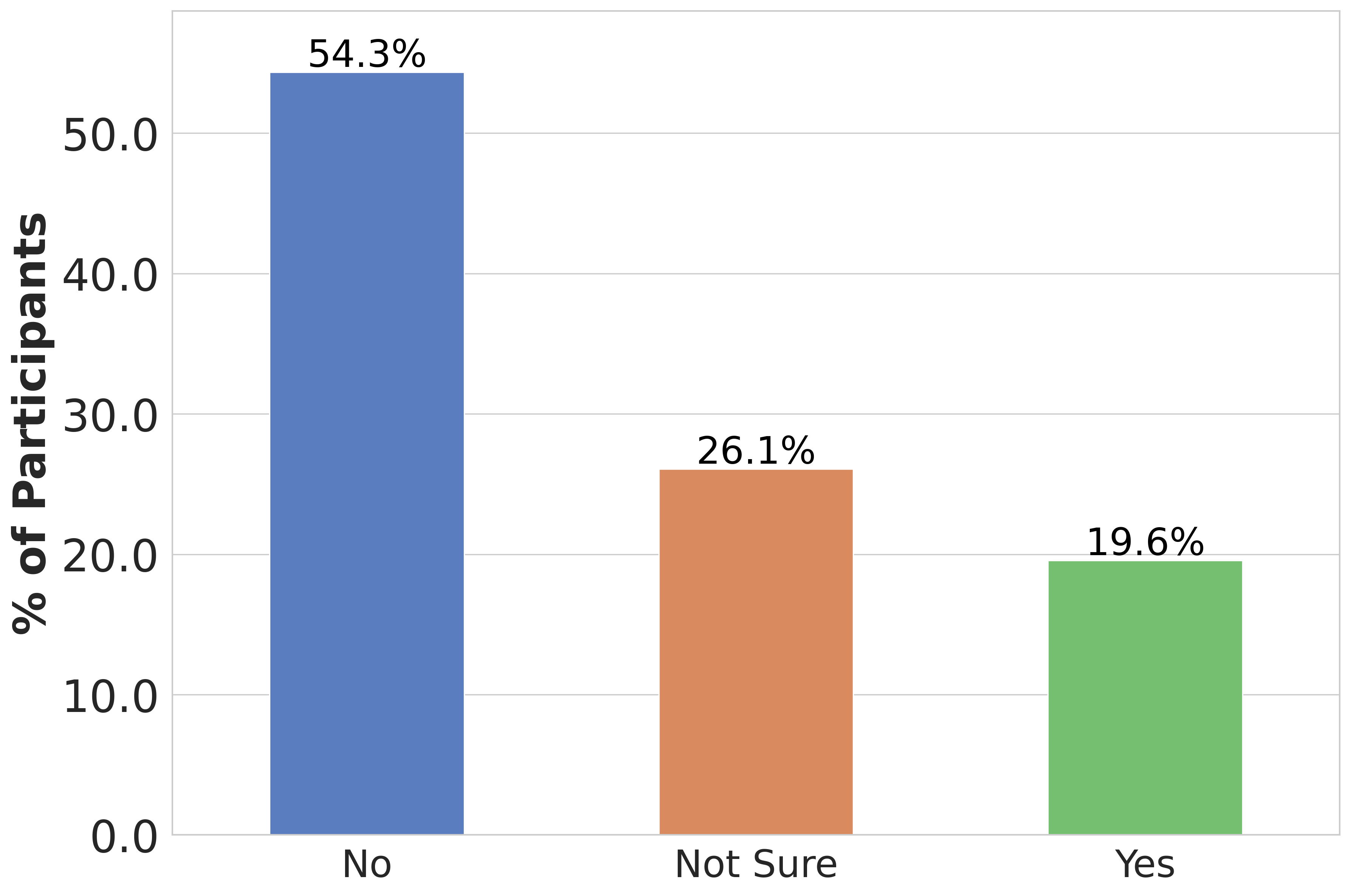}
        \caption{Influence of AI Tools on Health Decisions}
        \label{fig:ai_influence_health}
    \end{subfigure}

    \caption{Data visualizations for the survey responses (Part 1)}
\end{figure*}

\begin{figure*}[p]
    \centering
    % Fifth Row on a new page
    \begin{subfigure}{0.45\textwidth}
        \includegraphics[width=0.95\textwidth]{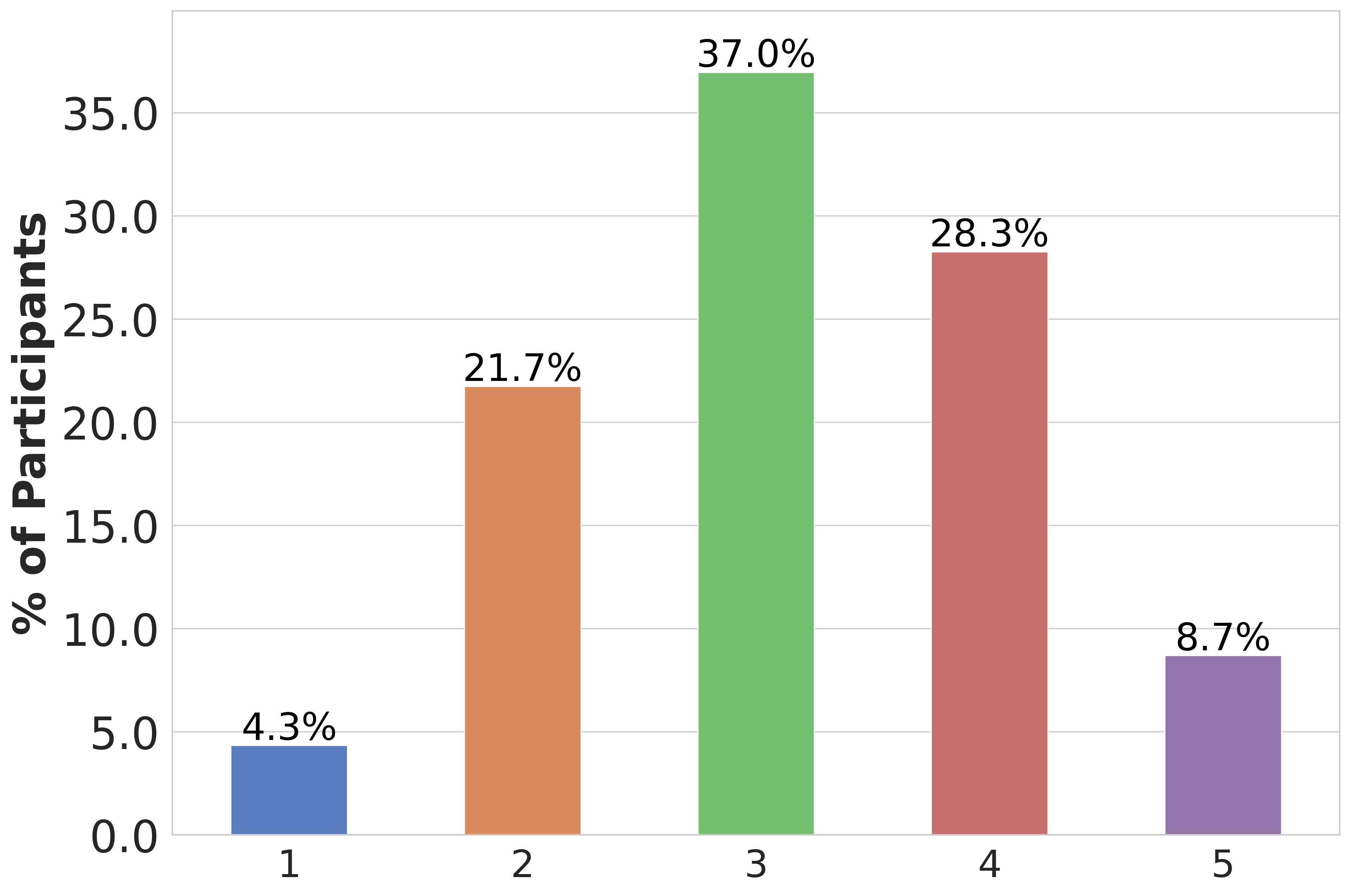}
        \caption{Rating AI Tools’ Accuracy and Relevance}
        \label{fig:rating_ai_tools}
    \end{subfigure}
    \hfill
    \begin{subfigure}{0.45\textwidth}
        \includegraphics[width=0.95\textwidth]{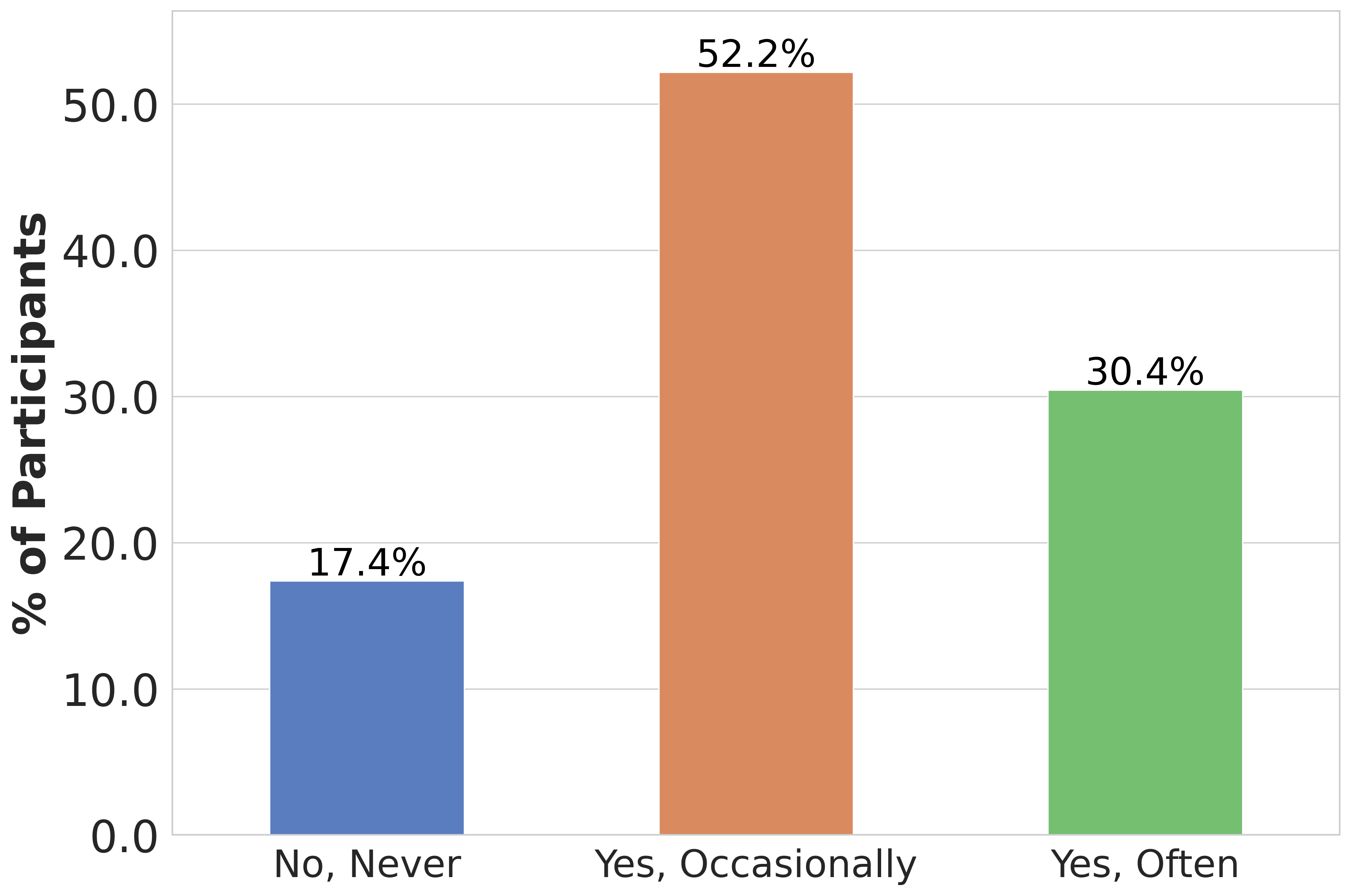}
        \caption{Verifying AI Tools’ Information with Other Sources}
        \label{fig:verify_ai_info}
    \end{subfigure}

    \caption{Data visualizations for the survey responses (Part 2)}
\end{figure*}

\begin{table}[H]
    \small
    \begin{tabular}{|p{6cm}|p{2cm}|}
        \hline
        \textbf{Web Technologies} & \textbf{Percentage} \\
        \hline
        Search Engines (Google, Bing, etc.) & 91.30 \\
        \hline
        Medical websites (WebMD, Mayo Clinic, etc.) & 47.83 \\
        \hline
        ChatGPT, Bard, Bing AI or such tools & 36.96 \\
        \hline
        Online medical forums and communities & 21.74 \\
        \hline
        Health-related apps & 10.87 \\
        \hline
    \end{tabular}
    \vspace{1em}
    \caption{\textbf{Web Technologies Used for Medical Information}}
    \label{table:web_technologies}
\end{table}

\begin{table}[H]
    \small
    \begin{tabular}{|p{6cm}|p{2cm}|}
        \hline
        \textbf{Patterns in Seeking Medical Information} & \textbf{Percentage} \\
        \hline
        Researching specific symptoms & 89.13 \\
        \hline
        Understanding treatment options & 69.57 \\
        \hline
        Reading about personal experiences & 43.48 \\
        \hline
        Comparing different sources & 32.61 \\
        \hline
        Seeking second opinions & 26.09 \\
        \hline
        Other & 2.17 \\
        \hline
    \end{tabular}
    \vspace{1em}
    \caption{\textbf{Patterns in Seeking Medical Information Online}}
    \label{table:seeking_medical_info}
\end{table}

\begin{table}[H]
    \small
    \begin{tabular}{|p{6cm}|p{2cm}|}
        \hline
        \textbf{Motivations for Using AI Tools} & \textbf{Percentage} \\
        \hline
        Quick answers to health-related questions & 63.04 \\
        \hline
        Convenience and accessibility & 47.83 \\
        \hline
        Seeking information outside of regular office hours & 23.91 \\
        \hline
        Privacy (no need to share personal details with a human) & 23.91 \\
        \hline
        Second opinions & 23.91 \\
        \hline
        To understand the hazards of treatment & 2.17 \\
        \hline
        Never used it for healthcare purpose & 2.17 \\
        \hline
    \end{tabular}
    \vspace{1em}
    \caption{\textbf{Motivations for Using AI Tools in Healthcare}}
    \label{table:motivations_healthcare_ai}
\end{table}

\begin{table}[H]
    \small
    \begin{tabular}{|p{6cm}|p{2cm}|}
        \hline
        \textbf{Challenges in Using AI Tools} & \textbf{Percentage} \\
        \hline
        Uncertainty about the source of information & 58.70 \\
        \hline
        Concerns about information accuracy & 52.17 \\
        \hline
        Lack of personalized advice & 34.78 \\
        \hline
        Difficulty in understanding medical jargon & 23.91 \\
        \hline
        Ambiguity in responses & 21.74 \\
        \hline
        Main issue with using ChatGPT & 2.17 \\
        \hline
        Not used for health care. Yet to explore & 4.34 \\
        \hline
    \end{tabular}
    \vspace{1em}
    \caption{\textbf{Challenges in Using AI Tools for Healthcare Information}}
    \label{table:challenges_healthcare_ai}
\end{table}

\begin{table}[H]
    \small
    \begin{tabular}{|p{6cm}|p{2cm}|}
        \hline
        \textbf{Interaction Preferences with AI Tools} & \textbf{Percentage} \\
        \hline
        Text-based chat & 82.61 \\
        \hline
        Visual representations (graphs, diagrams, etc.) & 41.30 \\
        \hline
        Voice-based interaction & 28.26 \\
        \hline
        Other & 4.34 \\
        \hline
    \end{tabular}
    \vspace{1em}
    \caption{\textbf{Interaction Preferences with AI Tools for Healthcare}}
    \label{table:interaction_preferences}
\end{table}